# PROGRAM WWZ: WAVELET ANALYSIS OF ASTRONOMICAL SIGNALS WITH IRREGULARLY SPACED ARGUMENTS

Ivan L. Andronov[1], Alevtina V. Naumova[2]

Department "High and Applied Mathematics", Odessa National Maritime University
Odessa, Ukraine
[1] il-a @ mail.ru, [2] lusik93 @ list.ru

ABSTRACT. A program WWZ is introduced, which realizes the wavelet analysis using an improved modification of the algorithm of the Morlet wavelet for a general case of irregularly spaced data, which is typical for the databases available in virtual observatories. Contrary to the well-known analogs, working with regularly spaced (equidistant in time) arguments, we have implemented an improved algorithm presented by Andronov, (1998KFNT...14..490A, 1999sss..conf...57A), which significantly increases the signal-to-noise ratio. The program has been used to study semi-regular pulsating variable stars (U Del et al.), but can be used for the analysis of signals of any nature.

One of the main directions of modern astrophysics is the study of variable stars. Specificity of astronomical observation is that in many cases the times at which the observation are obtained, are irregularly spaced. This is characteristic to collections of photographic negatives "Sky Patrol", available in a variety of observatories, as well as to more modern CCD sky surveys. Such observations are sometimes available through the Internet and included in national and international projects "Virtual Observatory". In Ukraine there are programs VIRGO ( Virtual X-ray and gamma-ray observatory) and UkrVO ( Ukrainian Virtual Observatory (Vavilova et al. 2012)).

Wavelet analysis is a method of investigating changes in the period of signals that are not strong periodic, but it is has sense to introduce "instantaneous" values of the "period" or "cycle length". In astronomy, such signals are called "quasi-periodic oscillations" (QPO) and are common to many types of variable stars - low-mass X-ray binaries, cataclysmic and symbiotic systems, semi-regular pulsating stars etc.

Mathematical foundations of the theory of wavelets are presented e.g. in the monographs by Daubechies (1992) and Vitjazev (2001) and reviews (e.g. Astafieva 1996). For the time series analysis, often is used the Morlet wavelet, which is based on a combination of the Fourier transform with the weight function being proportional to the Gaussian function. This conversion is included in the popular software, e.g. MatLab, SciLab. This algorithm allows to analyze only the signals, which are equally spaced in time, i.e. $t_k = t_i + (k-i)\Delta t$, where is $t_k$ is argument (typically time) of the $k^{th}$ observational point and $\Delta t$ is step (discretization in time). So for astronomical time series, Szatmary et al. (1994) have proposed to used simplified formulae, just changing integrals to sums on all observations.

As for the irregularly spaced time series, the basic functions do not satisfy conditions of orthogonality, the simplified formulas lead to a dependence of the 2D test function on the origin of the signal (i.e., its mean value), thus depriving the analysis of the physical meaning. Foster (1996) proposed to use the method of least squares with additional weights, and has introduced the test function «WWZ». A detailed analysis of the statistical properties of this and three other test functions was given by Andronov (1998, 1999), which is a special particular case of using arbitrary basic functions $f_\alpha(z)$ and the weight (filter) functions $w(z)$ (Andronov 1997).

The approximation may be written in a form:

$$x_c(t, t_0, P) = \sum_{\alpha=1}^{m} C_\alpha f_\alpha(z) \qquad (1)$$

Here $x_c$ is a smoothed value of a function which depends on the trial time $t$, on a "shift" $t_0$ and a "scale" (period) $P$; $z = (t-t_0)/P$, $C_\alpha(t_0, P)$ - the coefficients of expansion of the local signal on basic functions $f_\alpha(z)$, and $m$ - the order of the model.

For the „least squares extension" of the Morlet wavelet, $f_1(z) = 1$, $f_2(z) = \cos(2\pi z)$, $f_3(z) = \sin(2\pi z)$, $m = 3$. The weight function $w(z) = \exp(-cz^2)$ is the Gaussian function, where $c$ characterizes the width of the smoothing filter. When $c=0$, $w(z)=1$, and we have a global approximation of the sine wave. When $c \to \infty$, the filter function decays very quickly, and one obtains an asymptotic approximation of the local polynomial of degree 2, which corresponds to the method of running parabolae (Andronov, 1990, 1997). For applications, the most commonly used is an intermediate value $c=1/80 \approx 1/8\pi^2$. In the case of insufficient data, the values of $c$ may be decreased by a factor of 4 (corresponding to twice larger time resolution and twice smaller period resolution). A comparison of results of the wavelet analysis for different values of $c$ was presented also by Chinarova (2010) for the semi-regular variable RU And. Also, as a weighting function, one may use a local function $w(z)=(1-z^2)^2$ for $|z|<1$ (Andronov, 1990, 1997).

We had developed a program to evaluate few test functions: $r$ - semi-amplitude of oscillations, $S$ - the ratio of the weighted signal variance to the total variance (often denoted in other mathematical packages through $r^2$ - the square of the correlation coefficient between the observed and smoothed values, i.e. the same letter $r$ has another meaning), and WWZ - the "signal/noise" ratio.

As an illustration, we present the results obtained using our program for one of the semi-regular pulsating stars - U Delphini. A separate study of this star using various methods was presented by Andronov and Chinarova (2012). The original observations were published in the database of the French Association of Variable Star Observers "AFOEV", which is the precursor of modern databases of virtual observatories. The diagrams are shown in Fig. 1 for the model signal with a periodic modulation of the fundamental period and in Fig. 2 for U Delphini.

The program has been named WWZ, as one of the calculated test functions. Unlike previously worked analog (in the computer language Fortran, in text mode), the program WWZ was written in Object Pascal, and has a graphical interface and the ability to visualize both the original light curve and the test functions $F(t_0,P)$ (WWZ, $S$ and $r$), which depend on time (shift) $t_0$ and period (scale) $P$.

The graphic representation of the 2D dependence of the test function $F(t_0,P)$ is called "a wavelet map".

In addition, the program displays a "skeleton" - the time ($t_0$) dependence of the values of the periods corresponding to the local peaks at the wavelet map for a fixed $t_0$. Different colors show the peaks above and below the limit value $F_{max}(t_0)=\max_P(F(t_0,P))$, to facilitate the visual perception of the "skeleton". In the case of a constant period(s) "skeleton" should appear as a horizontal line(s). However, due to irregular spacing in time of the observations, the lines at the "skeleton" are apparently short or inclined. It also demonstrates the falsity of the respective peaks.

For convenience of the study of changes of the significance of different periods, we introduce a "cut mode ": the horizontal (time cross-section, $P$=const) and vertical (period cross-section, $t_0$=const), where the point $(t_0,P)$ is chosen by a mouse click. Corresponding two images are displayed below the "skeleton" and right from the wavelet map. For comparison, the two "wavelet periodograms" (also called "wavelet scalegrams") are shown - an average (shown in red) and "maximum" (for all time points within a "cut") . It should be noted, that, for small periods, the "maximum" value is many times greater than the average. This is due to the effective reduction of the number of points used for smoothing in the formula (1) and a corresponding increase in the scatter of the test function. Detailed mathematical description of the statistical properties was presented by Andronov (1998).

Because the relative widths of peaks at the scalegrams used in the wavelet analysis is substantially constant (in contrast to a classical periodogram, where the width of peaks is constant in frequency), instead of the period it is reasonable to use the logarithm of the period.

The range of periods is selected in accordance with a time variability of stars of corresponding types. For semiregular stars (including U Delphini), we used a range of $1.5 \leq \lg P \leq 4$. For stars with more rapid changes of brightness, the range is moved from seconds to about an hour.

The format of the input files for the described program: the light curve presented in two columns $(t_i, x_i)$, where $t_i$ = HJD- 2400000 - the time in Julian days and $x_i$ - stellar magnitude (or, alternatively, the intensity) of the object. Output file format used by 2. In the "full" format, the output is: the period $P$, the shift $t_0$ and test functions WWZ, $S$, $r$.

In the program, there are few restrictions for a case of "bad coverage" of the interval of smoothing by observations. E.g. there must be points after and before the shift $t_0$ and the effective number of observations should exceed 3. Otherwise the values of all the functions are set to zero. In this case, the wavelet map displays the white field.

For example, Fig. 2 shows that in the intervals, where there are no observations, the values are displayed only for large values of the period. That is, where the smoothing function is determined by the "distant" observations (before and after $t_0$). Obviously, such missing dots are not used in the weighted averaged wavelet map to obtain the corresponding periodogram.

In the short format, a special "variable length" algorithm: for each shift $t_0$ is determined the number $n_p$ of trial periods $P$, for which the value of the test function is nonzero. Then in the file are written $t_0$, $n_p$ and an array of values of the remaining test function. For sparing file size, only one of the test functions is chosen (namely WWZ).

The second output file contains the characterisstics of the skeleton, and the third one – the mean weighted wavelet periodogram ($\lg P$, <WWZ>, <$S$>, <$r$>)

**Conclusions**

The program for the wavelet analysis of the modified method of Morlet is introduced. The algorithm generalized to the general case of irregularly spaced (in time) observations was proposed by Andronov (1998). The program enables to calculate three different test functions with a meaning of: a) the "signal/noise" ration (WWZ); b) the square of the correlation ratio for the observed and smoothed values of the signal ($S$); c) the semi-amplitude $r$ of the smoothing sine function,

The program has the graphical interface and the ability to visualize the light curve and wavelet maps, cross-sections on both coordinates and the skeletons.

We developed the software used for the study of variable stars, and the results of this study will be used in the frame of the projects "Ukrainian Virtual Observatory" (UkrVO) (Vavilova et al., 2012) and "Inter-Longitude Astronomy' (Andronov et al., 2010).

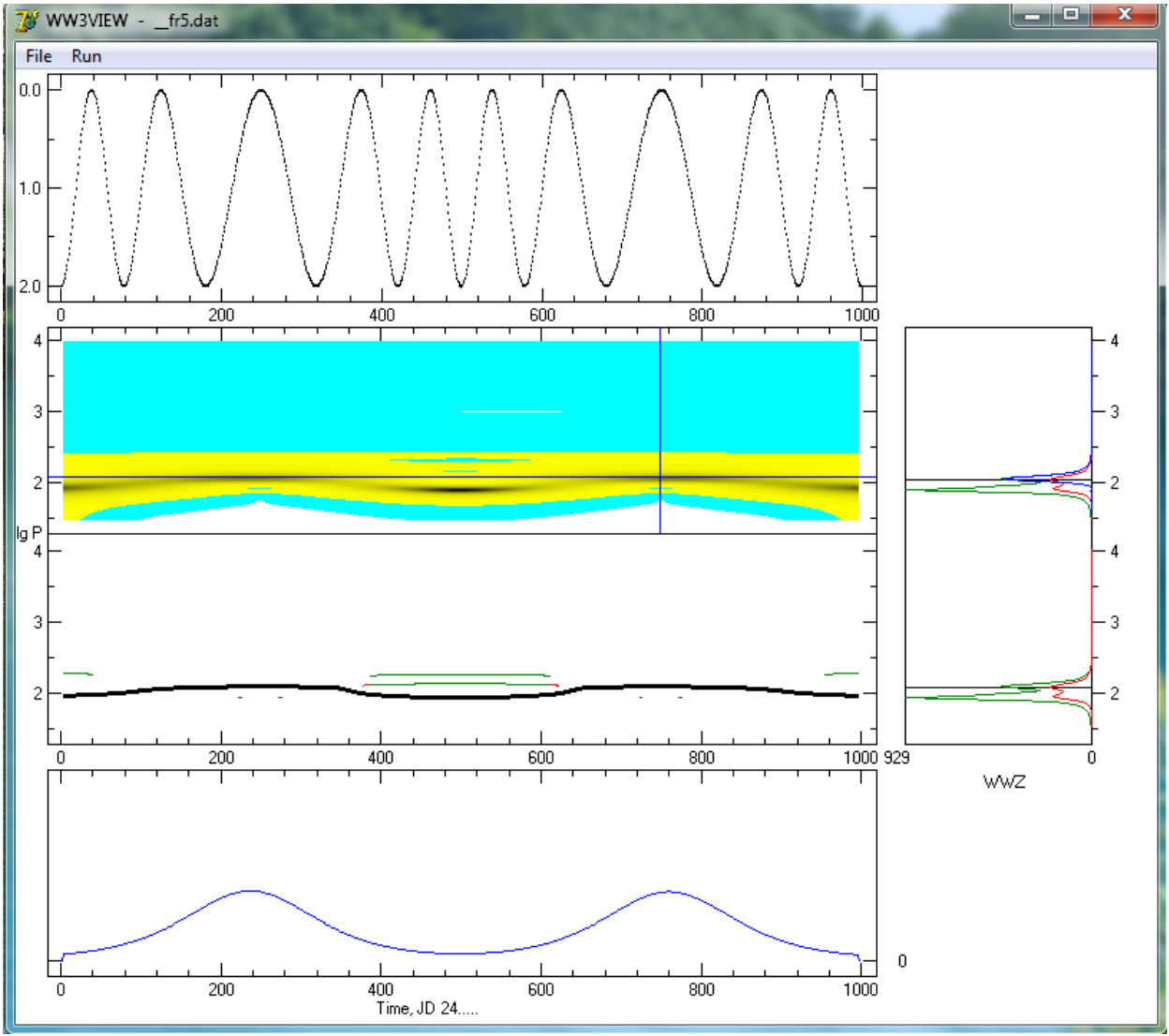

Figure 1. Screenshot of the program WWZ. From top to bottom: light curve of the model signal with rapidly changing period; the wavelet map (the dependence of the test function of time and of the logarithm of the period); the skeleton of the wavelet map (time dependence of the values of the period, the corresponding local peaks are highlighted in red peaks above a critical value equal to half of the maximum, and green - above the "threshold of detection", adopted at 3; thick black line indicates the position of the maximum for this shift ); the cross-section of the wavelet map in time for a fixed period. Right: blue marks a cross-section of the wavelet map at a fixed time (local periodogram), red shows the averaged periodogram, and green - the maximum value of the periodogram for a fixed period ant different shift.

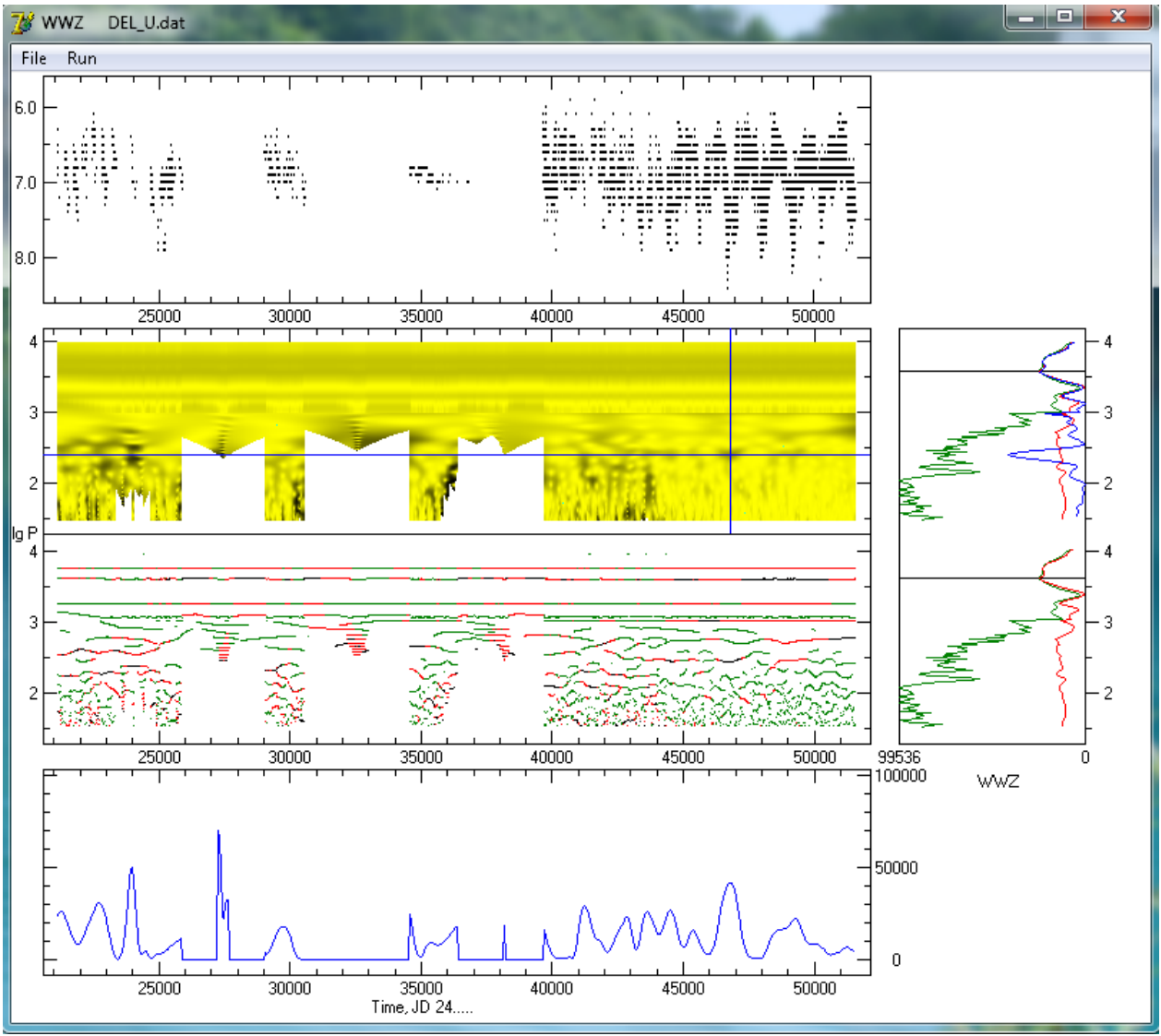

Fig.2. Screenshot WWZ for semiregular pulsating star U Delphini (similar to Fig.1). From top to bottom : light curve (the dependence of magnitude on the time in the Julian dates); the wavelet map; the skeleton of the wavelet map; the cross-section of the wavelet map in time for a fixed period. Right: blue markes a cross-section of the wavelet map at a fixed time (local periodogram), red shows the averaged periodogram, and green - the maximum value of the periodogram for a fixed period.